# Magnetic Nanoparticles for Power Absorption: optimizing size, shape and magnetic properties.

*M A Gonzalez-Fernandez*[1], *T Torres*[2], *M. Andrés-Vergés*[3], *R Costo*[1], *P de la Presa*[4], *C J Serna*[1], *M P Morales*[1], *C Marquina*[5], *M R Ibarra*[2,5], *G F Goya*[2,†].

[1]*Instituto de Ciencia de Materiales de Madrid, CSIC, Madrid, Spain*
[2]*Instituto de Nanociencia de Aragón, Universidad de Zaragoza, Zaragoza, Spain.*
[3]*Departamento de Química Orgánica e Inorgánica, Universidad de Extremadura, Badajoz, Spain*
[4]*Instituto de Magnetismo Aplicado, UCM-ADIF-CSIC, P.O. Box 155, 28230 Las Rozas, Madrid, Spain*
[5]*Instituto de Ciencia de Materiales de Aragón, CSIC, Zaragoza, Spain*

## Abstract

We present a study on the magnetic properties of naked and silica-coated $Fe_3O_4$ nanoparticles with sizes between 5 and 110 nm. Their efficiency as heating agents was assessed through specific power absorption (SPA) measurements as a function of particle size and shape. The results show a strong dependence of the SPA with the particle size, with a maximum around 30 nm, as expected for a Néel relaxation mechanism in single-domain particles. The $SiO_2$ shell thickness was found to play an important role in the SPA mechanism by hindering the heat outflow, thus decreasing the heating efficiency. It is concluded that a compromise between good heating efficiency and surface functionality for biomedical purposes can be attained by making the $SiO_2$ functional coating as thin as possible.

[†] Corresponding autor: goya@unizar.es

# I. Introduction

The interest in using core/shell magnetic nanoparticles (NPs) for biomedical and bioengineering applications has been increasing over the last few years. These magnetic NPs can be widely used for in-vitro as well as in-vivo applications [1, 2] such as magnetic biosensing [3], cell separation [4], and contrast enhancement in magnetic resonance imaging [5]. Further applications are being developed such as tissue repair, magnetic hyperthermia treatments [6], targeted drug delivery [7, 8] and labeling of cells [9]. The size, shape and biochemical coating of these nanoparticles are key attributes that must be controlled accurately [10]. For specific applications such as magnetic inductive hyperthermia (MIH), the rheological properties of the colloids and the efficiency for absorbing radiofrequency (RF) power on NPs also require optimization [11]. It is therefore important to characterize the magnetic properties so that the performance of the final pharmaceutical product can be evaluated.

Hyperthermia is a well known clinical protocol seeking to raise the temperature of a targeted body tissue above the physiologic level (c.a. 45°C or higher), usually for oncology applications. Based on the same rationale, the magnetic hyperthermia is a new technique developed to kill targeted cells by increasing the temperature of the intracellular medium using magnetic nanoparticles [12]. The amounts of NPs incorporated by a single living cell is of the order of few picograms, meaning that a relatively small number of NPs (~$10^3$ to $10^4$) have to be capable to rise the intracellular temperature by several degrees Celsius. The capability of a given material to generate heat from the magnetic coupling to an external alternate magnetic field is given by the SPA (Specific Power Absorption, also known as specific power losses SPL), which is the power absorbed per unit mass of magnetic nanoparticles. A great amount of effort has been delivered to the understanding of the major mechanisms that govern heat generated by power absorption and, as the main magnetic parameters are being better controlled, larger values of SPA are continuously being reported along the last years. [13]

In this paper we study the magnetic behavior of magnetite NPs in connection with their optimization as agents for hyperthermia treatments. For this purpose, magnetite NPs of different sizes and shapes were prepared; some of them were stable in water at pH = 7. To further increase the pH-range of stability, several samples were further coated with silica to obtain a core/shell structure. Silica has several advantages over organic coatings such as its high resistance against biodegradation [14]. Additionally, silica coating makes relatively easy to control the interactions among dispersed colloidal particles [15], so that the resulting system displays high stability under aqueous conditions in a very wide range of pH. The above properties make silica-coated nanoparticles excellent potential candidates for biomedical applications based on intravenous administration, since biodegradability, size and surface properties are known to be the key parameters determining the final distribution of NPs in living organisms.[16]

| *Sample* | *TEM NP size (nm)* | *Stabilizing agent* | *$SiO_2$ thickness (nm)* | *Isoelectric Point* | *Hydrodynamic size at pH 7 (nm)* |
|---|---|---|---|---|---|
| A | 24±5 | | - | 5 | 71 |
| B | 30±8 | $SO_4^{=}$ | - | 5 | 115 |
| C | 45±8 | | - | 5.3 | 156 |
| D | 42±6 | Citric acid | - | 2 | 124 |
| E | 5±1 | None | - | 7 | polydisperse |
| F | 110±9 | None | - | 6.7 | Not stable |
| BSi | 30±8 | | 1 | 2.5 | 150 |
| CSi | 45±8 | $SiO_2$ | 4.5 | 2.8 | 300 |
| FSi | 110±9 | | 15 | 1.25 | 423 |

**Table I.** *Summary table showing isoelectric points and hydrodynamic size of uncoated and coated samples.*

## II. Experimental

### A. Synthesis and coating of magnetic nanoparticles

The $Fe_3O_4$ nanoparticles were prepared by a direct method described elsewhere [17]. This method is based on the precipitation of an iron (II) salt ($FeSO_4$) in the presence of NaOH and a mild oxidant ($KNO_3$) at 90º C in a mixture of solvents water/ethanol. Particle size was controlled by changing the concentrations of the iron salt leading to magnetite nanoparticles with sizes from 24 nm to 45 nm and narrow size distributions (samples A, B and C, see Table I). The presence of ethanol in the medium not only control the speed of the reaction to produce cubic nanoparticles but also is responsible for the presence of sulphate ions $SO_4^{=}$ on their surface providing stability at pH 7. To assess the influence of particle shape, we followed the same synthesis route in water [17], resulting in spherical nanoparticles of 42 ± 6 nm size with narrow size dispersion. These particles were stabilized in water at pH 7 with citric acid [18] (Sample D).

The cubic nanoparticles having sizes of 30 nm and 45 nm were coated with silica following the Stöber method [19, 20]. A thin silica layer was deposited on their surface at a constant temperature of 20 ºC. The magnetite nanoparticles (30 mg) were added to a solution of 110 ml of 2-propanol that contained distilled water (12 ml) and ammonium hydroxide (1.5 ml). The solution was maintained in an ultrasonic bath for 1 h. Then, tetraethoxysilane (TEOS) was added to the solution and left in the ultrasonic bath for 6 h and 12 h depending on the required thickness of the silica layer. These samples were labeled as samples BSi and CSi, respectively. The solution was filtered, and the nanoparticles were washed with 2-propanol and dried at 20 ºC for 1 day. Then, they were re-dispersed in distilled water.

To perform a systematic study on the power absorption efficiency as a function of particle size, two additional samples were synthesized at the low and high ends of the series. First, small cubic magnetite nanoparticles (d ≈ 5 nm) were synthesized following Massart's method in the presence of ethanol (Sample E).[21] For obtaining larger $Fe_3O_4$ nanoparticles, hematite particles were first synthesized and then reduced to magnetite as described elsewhere [22], resulting in average particle size d = 110 nm (sample F). From this sample, core/shell $Fe_3O_4/SiO_2$ nanoparticles with the same magnetic core size were obtained (sample FSi) following the Stöber method mentioned above.

## B. Characterization

Particle size and shape were studied by transmission electron microscopy (TEM) using a 200 keV JEOL-2000 FXII microscope. TEM samples were prepared by placing one drop of a dilute suspension of magnetite nanoparticles in acetone on a carbon coated copper grid and allowing the solvent to evaporate slowly at room temperature. Colloidal properties of the samples were studied in a Zetasizer Nano™ from Malvern Instruments. The hydrodynamic size of the particles in suspensions was measured by photon correlation spectroscopy and the zeta potential was measured as a function of pH at 25º C, using $10^{-2}$ M $KNO_3$ as electrolyte and $HNO_3$ and KOH to vary the pH of the suspensions. Samples in powder were prepared and characterized magnetically at room temperature using a Vibrating Sample Magnetometer (VSM 9 MagLab 9 T, Oxford Instruments) at room temperature. Zero-field-cooled (ZFC) and field-cooled (FC) curves were measured using a Superconducting Quantum Interference Device (MPMS XL, Quantum Design), between 5 K and 280 K, with cooling field $H_{FC}$ = 100 Oe. Data were obtained by first cooling the sample from room temperature in zero applied field (ZFC process) to the basal temperature (5 K). Then a field was applied and the variation of magnetization was measured with increasing temperature up to T = 280 K. After the last point was measured, the sample was cooled again to the basal temperature keeping the same field (FC process); then the M vs. T data was measured for increasing temperatures. Low-temperature hysteresis loops (5-280 K) were obtained in applied fields up to 5 T. Specific Power Absorption (SPA) measurements were done using a commercial ac applicator (model DM100 by nB nanoscale Biomagnetics) working at 260 kHz and field amplitudes from up to 16 mT, and equipped with an adiabatic sample space (~0.5 ml) for measurements in liquid phase. Temperature data was taken using a fiber optic temperature probe (Reflex™, Neoptix) immune to rf environments.

## III. Results and Discussion

### A. Morphology and size distribution of nanoparticles

The statistics to calculate the mean particle size and distribution was performed from TEM image analysis, using sets of > 100 particles in each case. The general aspect and morphology of uncoated nanoparticles is illustrated in Figure 1 for samples B, D and F (see also Table I). It has been previously shown [17] that for this synthesis route the particle growth rate is determined by the iron salt concentration and water/ethanol ratio. This in turn seems to determine the final shape, since particles prepared with an excess of [$Fe^{2+}$] are spherical (samples with average size > 100 nm no studied in the present paper), while those prepared with an excess of [$OH^{-}$] are cubic. Samples A, B, C and D display very low size dispersion around their mean values. Colloidal suspensions of particles A, B and C were directly obtained by simple ultrasonic treatment of the powders leading to very stable ferrofluids at pH 7. Sulphate anions present at the particle surface seem to be responsible for the colloidal stability that provides a biocompatible character to the suspensions. However, sample D required further stabilization in water at pH 7 by adding citric acid [18].

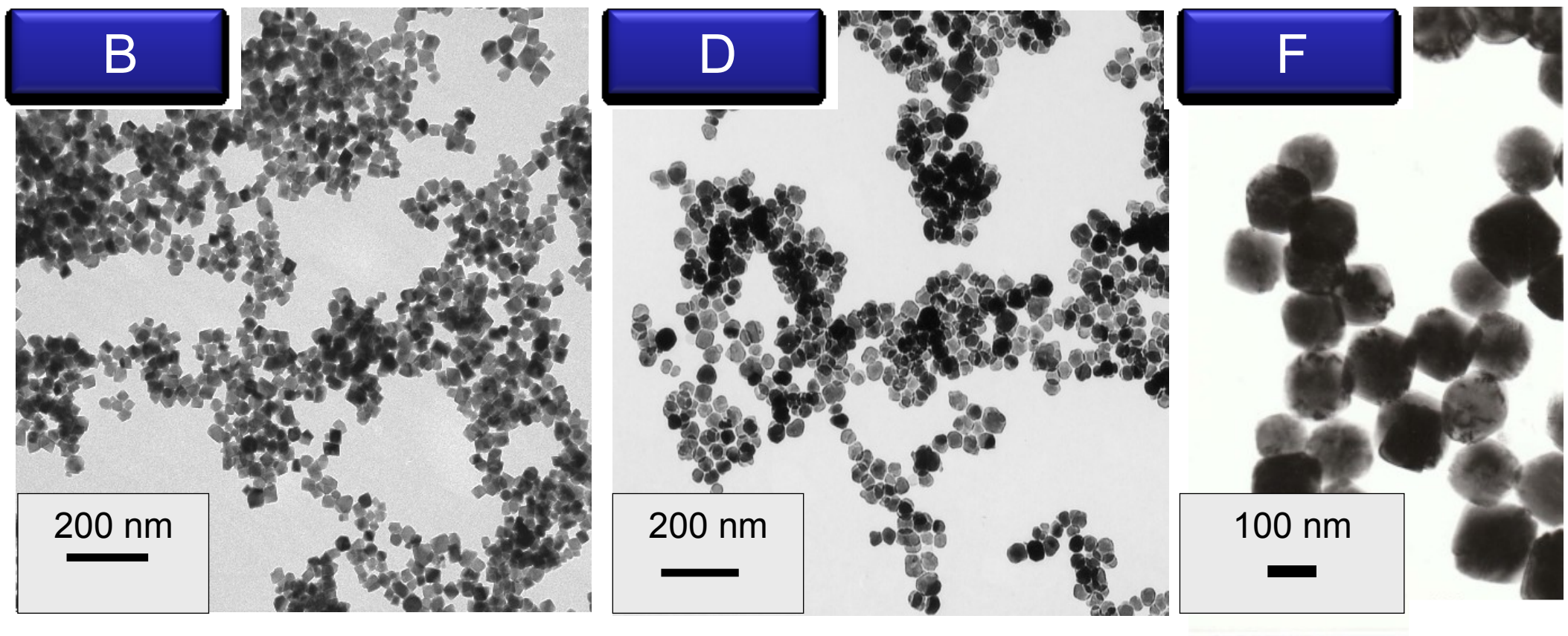

***Figure 1:*** *Low-resolution TEM images of samples B, D and F uncoated.*

The TEM image of sample CSi (Figure 2) shows clearly that the coating of the $SiO_2$ shell has uniform thickness and is complete around the magnetic cores. The thickness of the silica coating was extracted directly from the TEM images. Samples BSi (30 nm) and CSi (45 nm) were coated with a 1 nm and 5 nm silica layer, respectively. As explained in the experimental section, the control of the silica shell thickness was achieved through careful changes in the experimental parameters of the coating procedure, aiming to optimize the stability of the nanoparticles in water at high concentrations in a wide pH range.

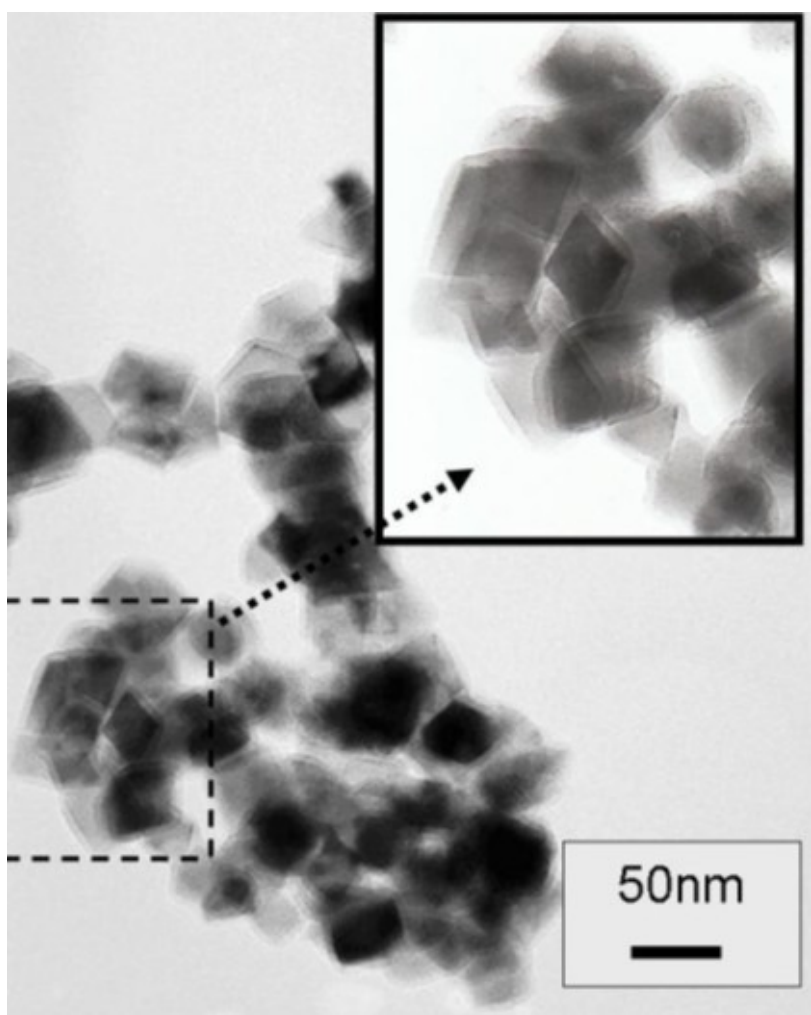

***Figure 2.*** *TEM image of sample CSi (45.5 nm size $Fe_3O_4/SiO_2$ NPs) coated with silica.*

## B. Colloidal properties

The colloidal stability of the magnetic nanoparticles at physiological values of pH and osmolality is a major issue if biomedical applications are considered. In the case of $Fe_3O_4$ particles, the zeta potential value reflects the resulting surface charge density that depends on the detailed oxide stoichiometry, degree of order at the particle surface and adsorbed molecules. In the present samples, negatively charged $SO_4^{=}$ groups are absorbed at the "naked" nanoparticle surface as a consequence of the synthesis method providing electrostatic repulsion and therefore stability at pH 7 but not at the physiological salinity [17]. It can be clearly seen (Figure 3) that the isoelectric points of samples B and C are shifted towards lower pH values respect to magnetite nanoparticles prepared by coprecipitation[23].

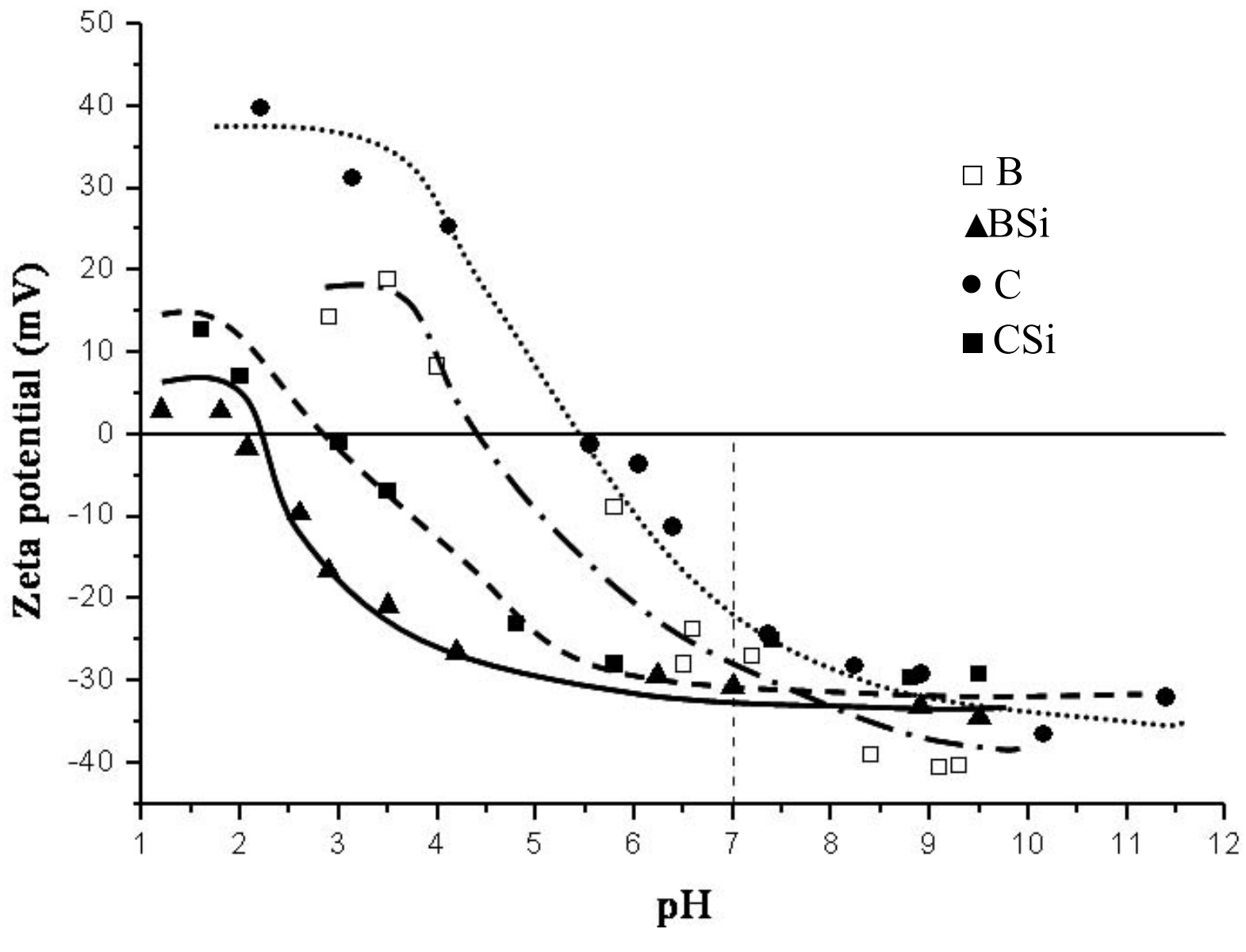

***Figure 3**. Zeta potential versus pH for 30 and 45 nm size coated and uncoated magnetite nanoparticles. The lines are guides to the eye.*

Some of the magnetic cores were further coated with a silica shell of different thickness since this kind of surface provides additional steric repulsion and can be easily functionalized with active groups like amine, carboxyl, aldehyde and thiol groups. The variation of the zeta potential as a function of pH is shown in Figure 3 for naked nanoparticles (samples B and C) as compared to silica coated ones (BSi and CSi respectively). Table I shows isoelectric point and the hydrodynamic size of coated and uncoated magnetite NPs studied in this work. The presence of silica on the surface of the nanoparticles in samples BSi and CSi results in a shift of the isoelectric point of these samples towards values near pH = 2. From the values of Z potential of the coated nanoparticles (see Figure 3), it is clear that the silica shell contributes to colloidal stability since they are stable in a wider pH range than the uncoated nanoparticles, as reflected in the lower isoelectric points of samples BSi and CSi [24]. These colloidal suspensions are stable in water at concentrations up to 5 mg/ml.

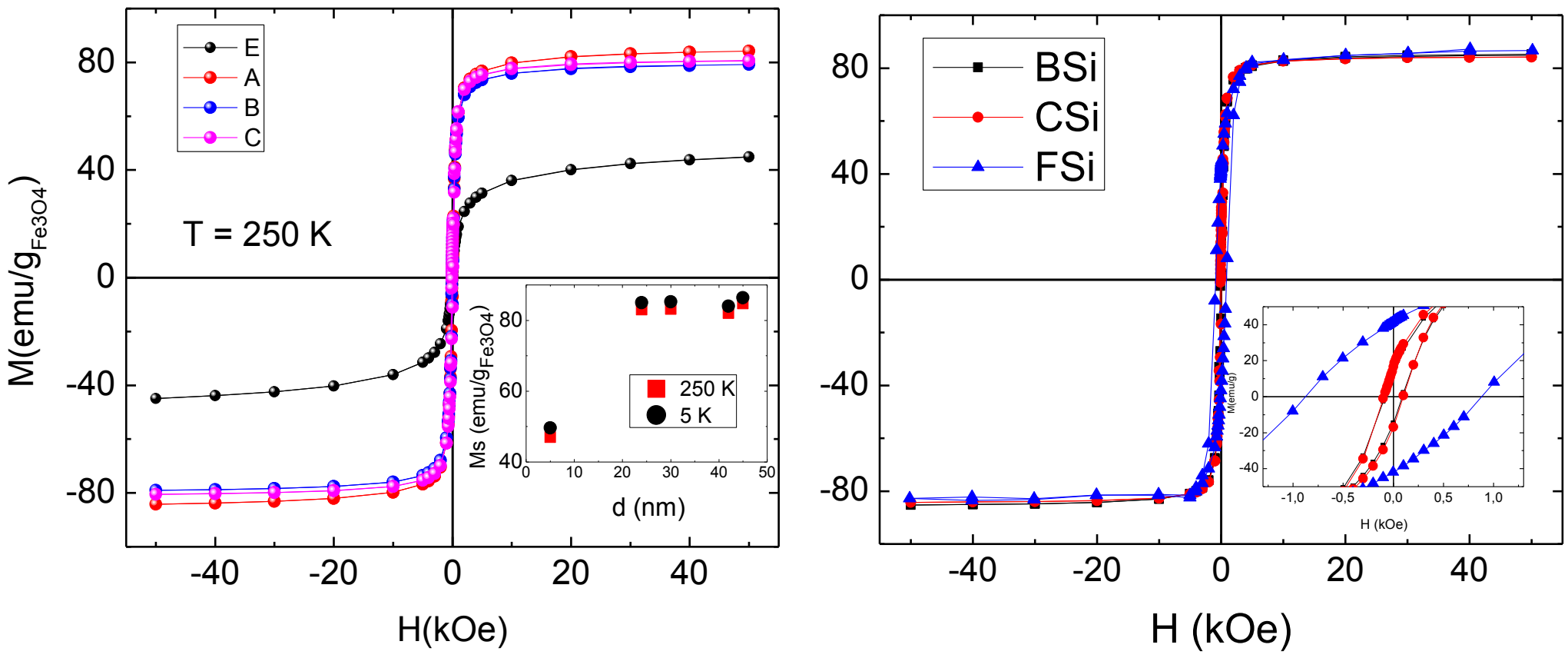

**Figure 4.** *Hysteresis M(H) curves for uncoated (left) and coated (right) magnetite nanoparticles taken at 250 K. Left inset: saturation magnetization as a function of average particle sizes taken at 5 and 250 K. Right inset: enlargement of the low-field region showing the different coercive fields for single- (BSi and CSi) and multi-domain (FSi) particles*

## C. Magnetic properties

In order to evaluate the correlation between basic magnetic parameters of the NPs and the efficiency for power absorption, we performed magnetization measurements as a function of temperature (ZFC-FC) and hysteresis loops at T = 5 K and 250 K. The coercivity values extracted from M(H) curves at T = 250 K (listed in Table II) were found to decrease somewhat with size, reflecting the effect of thermal energy on these (single domain) particles with $d < 50$ nm in the blocked state, which is to decrease the $H_C$ values for decreasing particle volume [25]. The saturation magnetization $M_S$ of these particles (Figure 4) showed essentially the same values within experimental accuracy (83-85 emu/g), in agreement with the fact that eventual contributions from surface disorder are expected to be similar. On the other hand, samples outside this narrow size window (i.e., samples E and F) have a distinct behavior. In the case of sample E having 5 nm, the value of $M_S$ = 47.7 emu/g was nearly half of the corresponding for the larger particles (see the inset of Figure 4). This reduction is a known effect from surface disorder, which increases in smaller particles (i.e., d < 10 nm) due to the increased surface/volume ratio[26]. On the other end of the series, the effect of the multi-domain structure in sample F (d = 110 nm) is reflected in the larger values of coercive field due to the domain wall displacement.

| Sample | d (nm) | t (nm) | Ms (emu/$g_{Fe3O4}$) T = 250 K | Mr/Ms T = 250 K | Hc (Oe) T = 250 K | Ms (emu/$g_{Fe3O4}$) T = 5 K | Mr/Ms T = 5 K | Hc (Oe) T = 5 K | SPA (W/g) |
|---|---|---|---|---|---|---|---|---|---|
| **A** | 24 | - | 83 | 0.08 | 44 | 85 | 0.3 | 311 | 137.4 |
| **B** | 30 | - | 83 | 0.12 | 64 | 85 | 0.3 | 309 | 83.6 |
| **C** | 45 | - | 85 | 0.14 | 75 | 87 | 0.41 | 242 | 62.7 |
| **D** | 42 | - | 82 | 0.15 | 79 | 88 | 0.32 | 169 | 11.7 |
| **E** | 5 | - | 48 | 0.004 | 4 | 50 | 0.27 | 334 | 3.17 |
| **F** | 110 | - | 77 | 0.3 | 170 | 88 | 0.33 | 250 | 1 |
| **BSi** | 30 | 1 | 85 | 0.17 | 86 | 89 | 0.04 | 165 | 81 |
| **CSi** | 45 | 4 | 86 | 0.19 | 93 | 87 | 0.15 | 100 | 45 |
| **FSi** | 110 | 15 | 80 | 0.5 | 890 | 86 | 0.6 | 2014 | 1.74 |

**Table II:** Magnetic parameters of $Fe_3O_4$ nanoparticles of different sizes and $SiO_2$ thickness.

Magnetization curves as a function of temperature taken in zero-field and field cooling modes for samples A and B (**Figure 5**) show that the NPs remain blocked up to room temperature, since no maximum in the ZFC is observed. For temperatures below ≈50 K both ZFC curves show a sudden decrease, usually observed for magnetite nanoparticles having sizes between 20 and 50 nm [17] and associated to a decreased Verwey transition of $Fe_3O_4$, located at $T_V$ = 125 K in the bulk material [27]. It has been previously reported that for particles having d < 50 nm the Verwey transition shifts to lower temperatures with decreasing particle size, and is no longer observable for d ≤ 10 nm.[26] In agreement with the smaller size of sample E, the corresponding ZFC-FC data shows the maximum in the ZFC curve at $T_B$ = 104.4 K (see inset of **Figure 5**)

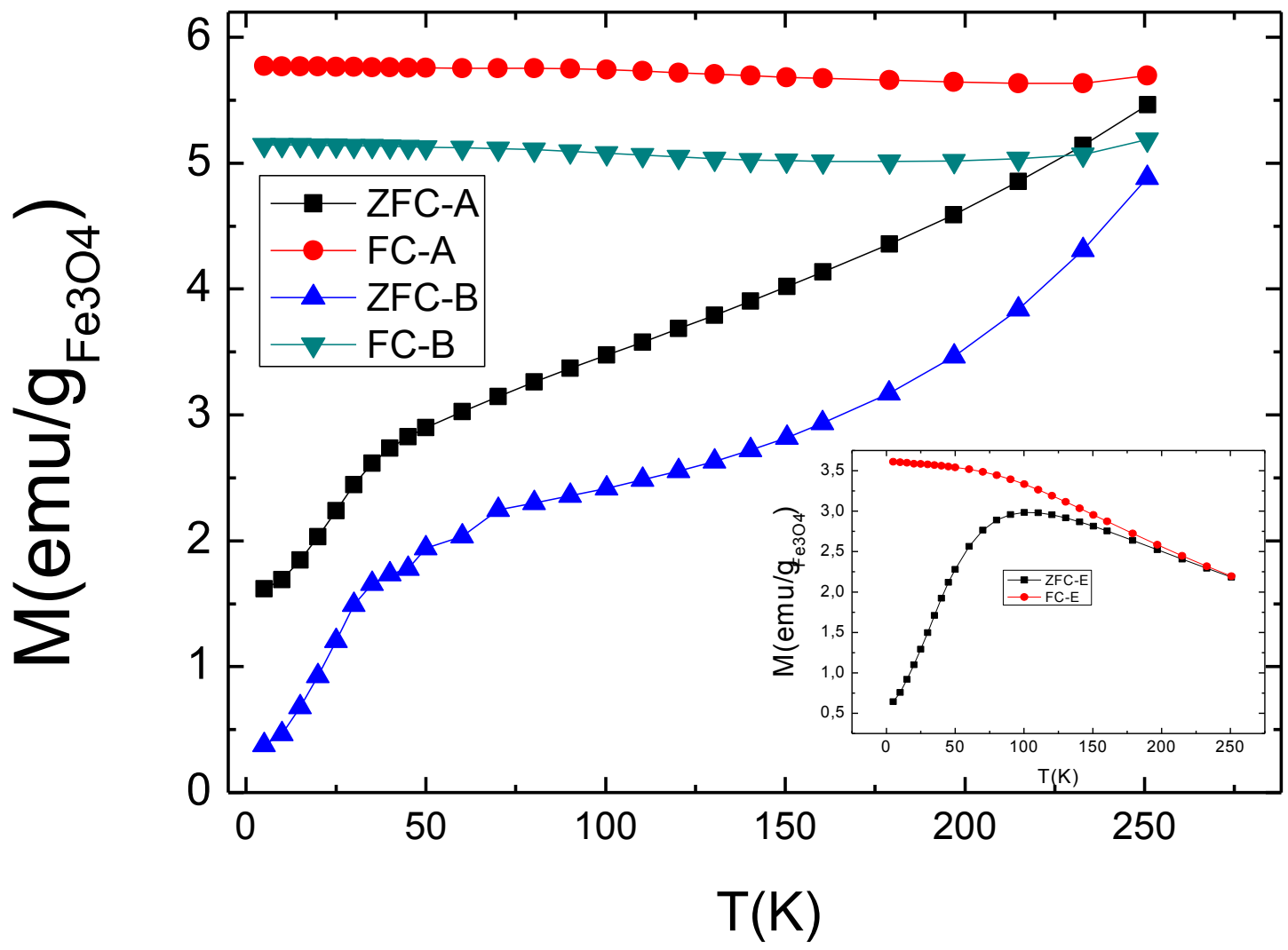

**Figure 5:** *ZFC-FC curves for samples A and B uncoated magnetite nanoparticles taken from 5 to 250 K. Inset: ZFC-FC curves for sample E show the $T_B$~104.4 K.*

## D. Effect of size on SPA

In order to assess the influence of physical parameters on heat generation, we measured the SPA as a function of particle size spanning the 20 to 45 nm size range. For completeness, two additional samples having average size values of 5 nm and 110 nm were also studied. The heating efficiency of the colloids was measured from the temperature increase (ΔT) of a given mass of the constituent nanoparticles ($m_{NP}$) diluted in a mass of liquid carrier ($m_{LIQ}$) during the time interval ($\Delta t$) of the experiment. The expression for power absorption $P$ per unit mass of the magnetic material is given by:

$$\Pi = \frac{P}{m_{NP}} = \frac{m_{LIQ} c_{LIQ} + m_{NP} c_{NP}}{m_{NP}} \left( \frac{\Delta T}{\Delta t} \right) \qquad (1)$$

where $c_{LIQ}$ and $c_{NP}$ are the specific heat capacities of the liquid carrier and the nanoparticles, respectively. Since the concentrations of MNPs are usually in the range of 1% wt. or less, we can approximate (1) by

$$\Pi = \frac{c_{LIQ} \delta_{LIQ}}{\phi} \left( \frac{\Delta T}{\Delta t} \right) \qquad (2)$$

where $\delta_{LIQ}$ and $\phi$ are the density of the liquid and the weight concentration of the MNPs in the colloid, respectively.

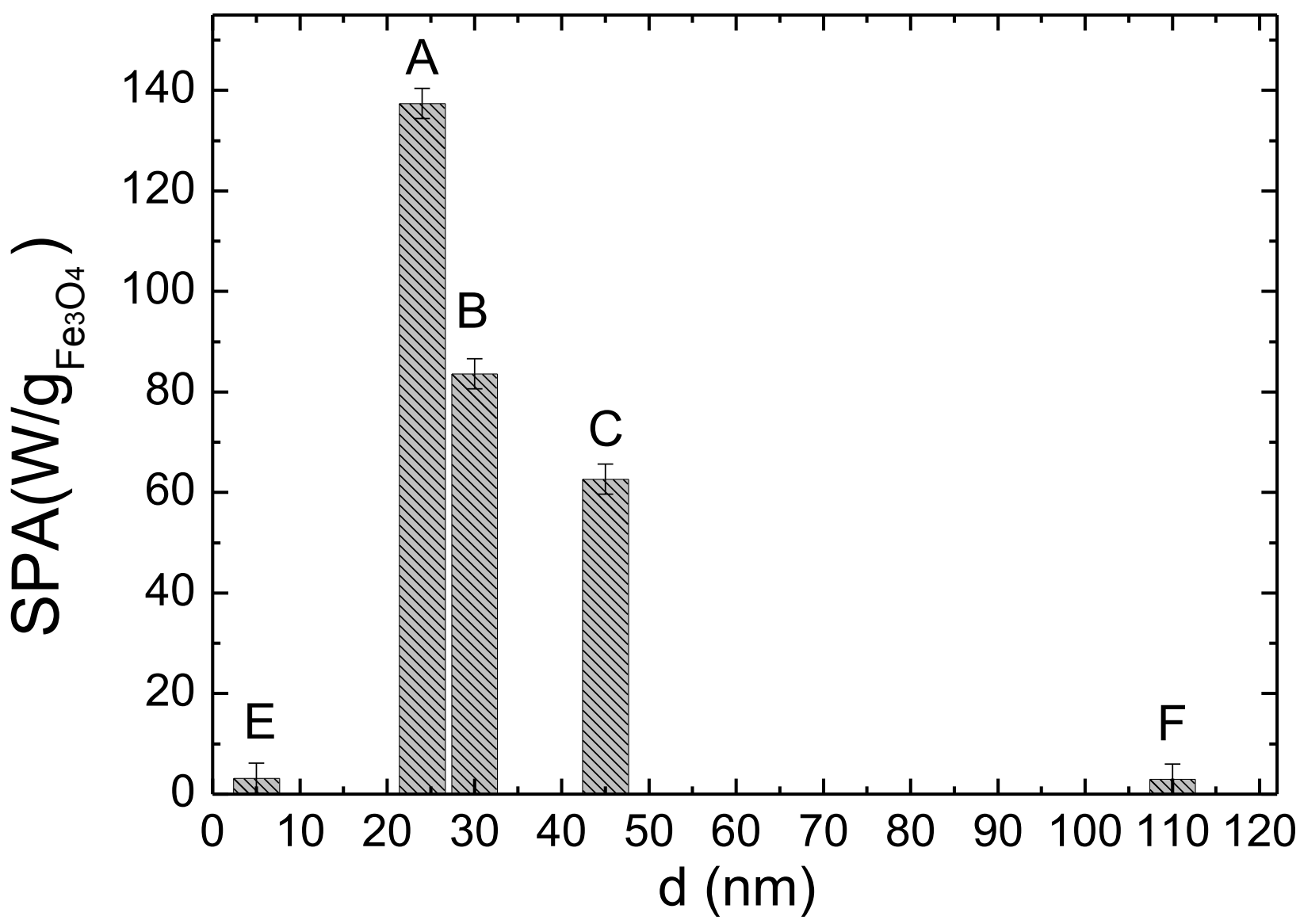

**Figure 6:** *Specific Power Absorption of uncoated magnetite samples with different nanoparticle size.*

The SPA values measured according to the experimental section are listed in Table II in W/g of magnetite and plotted as a function of particle size in Figure 6. It can be observed that SPA values are maximum for sample A with d = 24 nm and, for values below and above this maximum, the values decrease abruptly. This is in agreement with the expected behavior for Néel relaxation-based mechanism as first proposed by Rosensweig [28] and experimentally found in many colloidal systems [11, 17, 29].

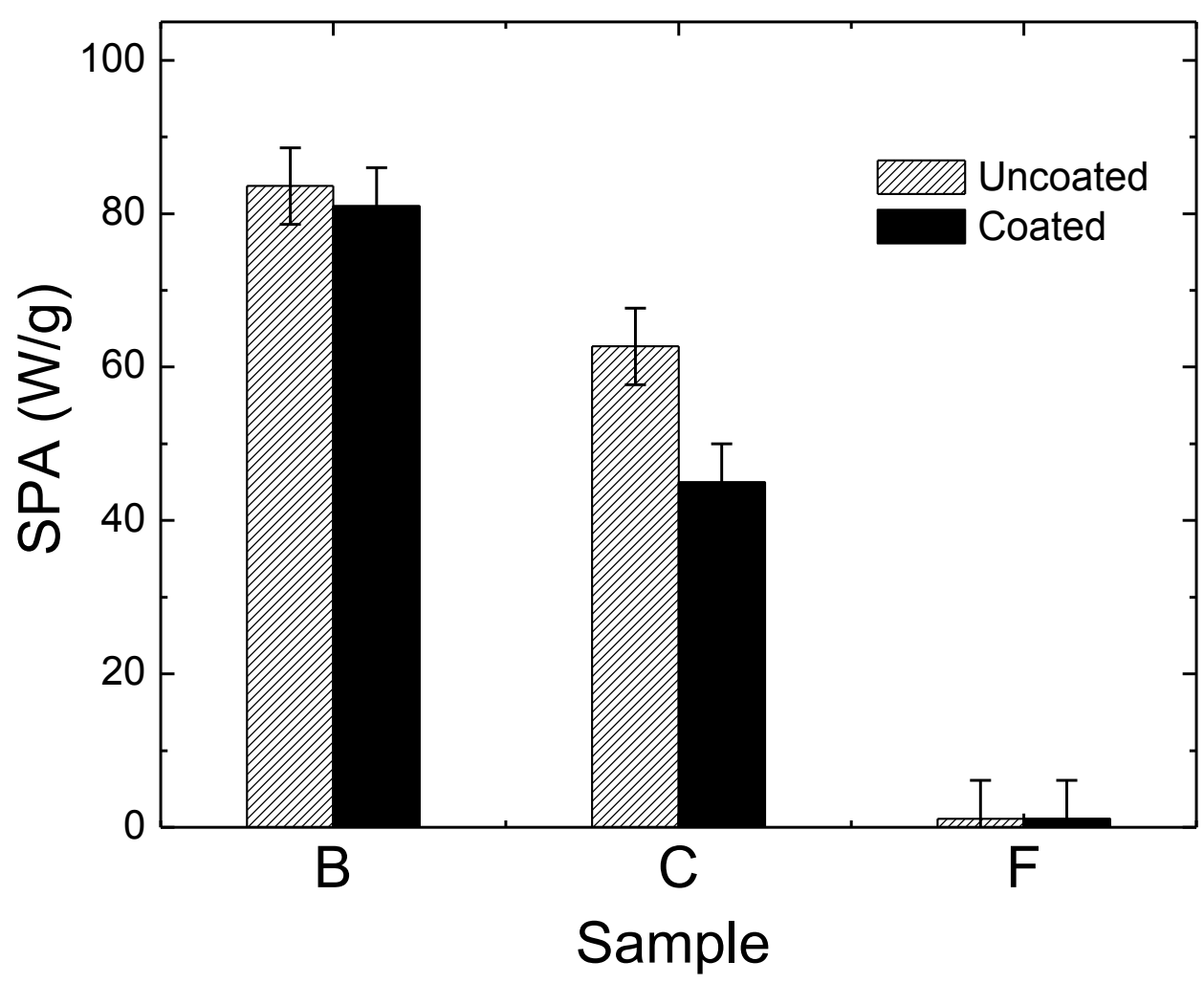

**Figure 7:***SPA of single domain (B and C) and multidomain (F) nanoparticles. The shaded bars correspond to uncoated (B, C and F) samples, and filled bars represent the silica-coated (BSi, CSi and FSi) samples.*

The Néel relaxation mechanism is strongly dependent on particle volume through the term $\tau \sim \exp\{KV/k_BT\}$, so that the optimum size of nanoparticles for hyperthermia should be located within a narrow size window (lower than the single-domain critical size for a given material). In addition to the drop of SPA values with increasing size within the single-domain region, it is worth to notice that SPA becomes essentially null for multi-domain particles (i.e., sample F). This small value for F sample suggests that domain wall displacements contribution to power absorption is much smaller than the Néel relaxation process in single domain particles at these frequencies. This observation is supported by previous measurements (not shown) performed on well-crystallized, multi-domain particles of $Fe_3O_4$ with 1 μm size and bulk material, which showed small SPA values of 0.2 and 0.3 W/g, respectively. Indeed, the maximum SPA value measured for bulk magnetite at $f$ =260 kHz and B = 100 Oe, which is mainly due to losses from domain wall displacements, was about 0.5 W/g.

### E. Effect of silica coating on SPA

To evaluate the effect of $SiO_2$ shell on the heat release process, we have determined the SPA of the silica-coated magnetite nanoparticles of sizes 30, 45 and 110 nm, labeled BSi, CSi and FSi respectively, and compared to the corresponding 'naked' particles (samples B, C and F). We recall the fact that all of these $SiO_2$-coated samples are composed of the same magnetic cores than the corresponding uncoated nanoparticles. Therefore the magnetic coupling and power absorption efficiency should be the same for the corresponding coated and uncoated samples with the same particle size. The comparative results of SPA measurements between naked and coated NPs are shown in Figure 7. In addition to the observed decrease of SPA with increasing particle size, already discussed, the coated NPs display smaller SPA values than the corresponding uncoated ones with similar sizes. Moreover, the largest reduction is observed for CSi sample, which has a thicker $SiO_2$ shell. One possible explanation for this difference could be related to the insulating nature of the silica coating, which could be shielding the heat from the magnetic nanoparticle. This indicates that the silica coating of the nanoparticles should be designed to be the minimum necessary to keep the nanoparticles stable in water. In the case of samples F and FSi, composed of the same magnetic cores with average size 110 nm, the SPA corresponds to domain wall losses since for this size range the NPs have a multi-domain structure [30] and the power losses are due to domain wall displacements.

### F. Effect of particle shape on SPA

Samples of 45 nm were obtained with cubic and spherical shapes (samples C and D, respectively) to explore the effects of anisotropy on the SPA. These samples have been characterized magnetically at different temperatures in powder form. The hysteresis loops of both samples measured at room temperature show similar $H_C$ values of 70-75 Oe. These values are consistent with single-domain particles near the blocking temperatures (which is above 250 K). However, at low temperatures where particles are fully blocked (T = 5 K) sample C displays a larger $H_C$ value, indicating a source of larger magnetic anisotropy (see Table II). We believe that this difference is likely to be originated in stronger shape anisotropy due to its cubic form, as observed from TEM images (*Figure 1*). Since both samples are of the same $Fe_3O_4$ crystals, magnetocrystalline anisotropy does not appear to be the source of this anisotropy difference. Accordingly, **Error! Reference source not found.** the power absorption values in the anisotropic nanoparticles (sample C), considerably larger than the corresponding for the spherical particles (sample D) with the same average size, indicate that shape anisotropy differences originated from particle morphology have to be considered in order to optimize specific power absorption values for hyperthermia applications.

## IV. Conclusions

We have succeeded in producing highly stable magnetic NPs with excellent control of particle size between ~20-50 nm. This control of final size allowed us to match the narrow size window to reach maximum power absorption efficiency for hyperthermia application. The maximum SPA values correspond to sizes around 30 nm, and falls down rapidly for values that differ in few nanometers. We have also observed that, in addition to size value, the magnetic anisotropy of the nanoparticles is relevant to obtain higher values of SPA. Coating the particles with an insulating $SiO_2$ shell resulted in lower SPA values due to changes in heat propagation out of the particles. This result indicates that the surface functionalization of the particles designed for heating therapies in biomedicine should be kept to a minimum.

## V. Acknowledgments

The authors acknowledge the financial support from Diputación General de Aragón, Comunidad de Madrid (S-0505/MAT/0194), and Ministerio de Ciencia e Innovación (MAT2005-03179 and MAT2008-01489). GFG and PP acknowledge support from the Spanish MEC through the Ramon y Cajal program.